\documentclass{article}

\usepackage{amsmath}
\usepackage{amsfonts}
\usepackage{amssymb}
\usepackage{amsthm}
\usepackage{enumitem}
\usepackage{float}
\usepackage{graphicx}
\usepackage[ruled, vlined]{algorithm2e}
\usepackage[margin = 1.5in]{geometry}
\restylefloat{table}
\usepackage{url}
\usepackage{todonotes}

\theoremstyle{definition}

\theoremstyle{plain}

\numberwithin{equation}{section}

\begin{document}

\title{Attacks on Image Encryption Schemes for Privacy-Preserving Deep Neural Networks}
\author{Alex Habeen Chang, Benjamin M. Case}
\date{\today}
\maketitle

\begin{abstract}

Privacy preserving machine learning is an active area of research usually relying on techniques such as homomorphic encryption or secure multiparty computation.  Recent novel encryption techniques for performing machine learning using deep neural nets on images have recently been proposed by Tanaka \cite{TanakaPublished} and Sirichotedumrong, Kinoshita, and Kiya \cite{SKK}. We present new chosen-plaintext and ciphertext-only attacks against both of these proposed image encryption schemes and demonstrate the attacks' effectiveness on several examples.\footnote{https://github.com/ahchang98/image-encryption-scheme-attacks}


\end{abstract}

\section{Introduction}

Secure computation is a major theme in modern cryptography with many different approaches to the general problem of how to encrypt data securely while still being able to use it for some form of computation.  Methods for secure computation include homomorphic encryption, multi-party computation, zero-knowledge proofs, functional encryption, and program obfuscation \cite{privacypreservingcomp}. A particularly interesting application area is to enable machine learning to be done on data while it is secured by encryption. Some work in this direction of implementing machine learning and deep neural nets using homomorphic encryption has been done as in \cite{boemer2019ngraph, PrivateAI, gilad2016cryptonets, lou2019glyph}.


Other novel techniques for performing machine learning using deep neural nets (DNNs) on images have recently been proposed; namely, the Tanaka scheme in \cite{TanakaPublished} and the Sirichotedumrong, Kinoshita, and Kiya (SKK) scheme in \cite{SKK}. The Tanaka scheme relies on performing deterministic encryption with the same key for all the images used for training and querying the DNN. The SKK scheme supports using different encryption keys for different images, and the encryption scheme preserves local properties of the images so that the machine learning can still be carried out.

We are able to break the Tanaka scheme using a chosen-plaintext attack.  This is not that surprising, since it is a result in theoretical cryptography that a deterministic encryption scheme can never satisfy the property of being CPA secure \cite{KatzLindell}. Tanaka himself admits in \cite{Madono} that ``scrambled images are not exactly encrypted although the perceptual information can be hidden" and ``effective algorithms for reconstructing original images from block-wise scrambled images are still available, to the best of our knowledge." So it is acknowledged that such attacks could happen, but we are not aware of other papers that have clearly demonstrated any attacks. 

We are also able break the variant of the SKK scheme that uses the same encryption key for each image using a chosen-plaintext attack. For the variant of the SKK scheme that uses a different encryption key for each image, a chosen-plaintext attack is unlikely, but we are instead able to develop ciphertext-only attacks that recover much of the original image. These attacks work by exploiting the fact that, in this scheme, much of the encryption is done with respect to each pixel without their movement. It is then easy to unscramble the color components at each pixel such that the gradient magnitude at each pixel is minimized to a certain extent. Indeed, in realistic input images it is commonly the case that, as we observe, the gradient magnitude at each pixel is close to minimal, due to the gradually changing color component values. The SKK encryption scheme is also used in the related paper \cite{SMKK} and our attacks apply there as well.

\section{Proposed Image Encryption Schemes}

We will represent an image \( I \) in this paper as a matrix of size \( U \times V \), where each element of the matrix is a pixel \( p = (u, v) \). Each pixel has three color channels red, green, and blue, labeled \( p_R \), \( p_G \), and \( p_B \), respectively, each of which is stored as an \( L \)-bit integer (i.e. as a value in \( [0, 2^L - 1] \)). For example, for an 8-bit image \( L = 8 \).

We describe to the best of our knowledge below the two proposed image encryption schemes that we attack later.

\subsection{Tanaka Scheme}

We present a version of the Tanaka scheme that we believe is equivalent to the one presented in the original paper \cite{TanakaPublished}.

Suppose an \( L \)-bit full color image \( I \) and secret encryption key \( K \) are given as input; the Tanaka scheme proceeds as follows:
\begin{enumerate}

\item \( I \) is divided into blocks of predetermined size \( M \times M \) (\( M = 4 \) in the experiments presented in \cite{TanakaPublished}.) Consider such a block \( P \).

\item With respect to the secret encryption key \( K \), the color components within \( P \) are shuffled. More precisely, the procedure is this: choose \( p \) and \( p' \) as pixels in \( P \), and choose \( C \) and \( C' \) in \( \{ R, G, B \} \). Then swap \( p_C \) and \( p'_{C'} \); repeat this procedure as desired. The manner in which the pixels are shuffled is the same for all blocks \( P \).

\item Optionally, with respect to the secret encryption key \( K \), the values of pseudorandomly selected color components within \( P \) are reversed (e.g. a color component \( p_B \) of a pixel \( p \in P \) originally with value 57 gets \( 255 - 57 = 198 \), where \( 255 = 11111111_2 \) is the max value for 8-bit images). The manner in which the values of the pixels are reversed is the same for all blocks \( P \).

\end{enumerate}

\begin{figure}[H]
\centering
\includegraphics[width = 96px, height = 96px]{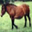}
\qquad
\includegraphics[width = 96px, height = 96px]{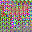}
\caption{An 8-bit input image (original size \( 32 \times 32 \)) and an encryption using the Tanaka scheme with block size \( M = 4 \) and optional reversing.}
\end{figure}

(We remark that, although the original paper states ``each block is split to the upper 4-bit and the lower 4-bit images", the authors of this paper did not find any useful interpretation of this step when comparing with the code provided in \cite{TanakaPublished}.)

We believe that Tanaka trained the networks used for validation accuracy testing using sets of pictures encrypted with the same key. Indeed, even if Tanaka trained using sets of pictures encrypted with different keys, \cite{SKK} experimentally shows that image classification accuracy when testing DNN models trained using these images is very low compared to DNN models trained using sets of pictures encrypted using other schemes. So the Tanaka scheme using different encryption keys is potentially undesirable for the wanted applications in machine learning and deep neural networks.

\subsection{Sirichotedumrong, Kinoshita, and Kiya (SKK) Scheme}

We present a version of the SKK scheme that we believe is equivalent to the one presented in the original paper \cite{SKK}.

Suppose an \( L \)-bit full color image \( I \) and secret encryption key \( K \) are given as input; the SKK scheme proceeds as follows:
\begin{enumerate}

\item \( I \) is divided into individual pixels. Consider such a pixel \( p \).

\item \emph{The Negative-Positive Transformation}: With respect to the secret encryption key \( K \), a pseudorandom bit \( x_R \in \{ 0, 1 \} \) dependent on the key is generated and the pixel value \( p_R \) is modified using
\[
p_R = 
\begin{cases}
p_R & (x_R = 0)\\
p_R \oplus (2^L - 1) & (x_R = 1)
\end{cases}
\]
(where XOR with \( 2^L - 1 = 111 \ldots 111_2 \) is to flip the bits of the value \( p_R \)). The same is done for \( p_G \) and \( p_B \): that is, pseudorandom bits \( x_G, x_B \in \{ 0, 1 \} \) are generated from the key, not necessarily equal to \( x_R \), and the pixel values of \( p_G \) and \( p_B \) are modified in the same way. The values of \( x_R \), \( x_G \), and \( x_B \) are not necessarily the same for all pixels \( p \).

\item Optionally, with respect to the secret encryption key \( K \), a pseudorandom integer \( x_S \in [0, 5] \) is generated and the values of \( p_R \), \( p_G \), and \( p_B \) are shuffled, where each of the six possible integers corresponds to a unique possible permutation. The value of \( x_S \) is not necessarily the same for all pixels \( p \).

\end{enumerate}

\begin{figure}[H]
\centering
\includegraphics[width = 96px, height = 96px]{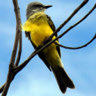}
\qquad
\includegraphics[width = 96px, height = 96px]{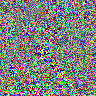}
\caption{An 8-bit input image (original size \( 96 \times 96 \)) and an encryption using the Sirichotedumrong, Kinoshita, and Kiya scheme with optional shuffling.}
\end{figure}

The experiments in \cite{SKK} cite results both using the same encryption key and different encryption keys to encrypt many images. 

\section{Attacks}

We present below our attacks on the two proposed image encryption schemes described above.

\subsection{Attacking the Tanaka Scheme}

For the Tanaka scheme we present a \emph{full} chosen-plaintext attack: that is, we are certainly able to recover the original image with all its colors and features. We have no full nor partial ciphertext-only attack against the Tanaka scheme at this time.

Being able to query the encryption scheme up to \( O(M^2) \) times with chosen plaintexts to obtain the corresponding ciphertexts, we wish to fully decrypt a given ciphertext \( \texttt{Enc}(I) \) of the original 8-bit image \( I \).  To do this, recall that the manner in which the pixels are shuffled and in which the values of the pixels are reversed is the same for all blocks. Recall also that this scheme uses the same encryption key for each input image, such that the block operations are the same across different ciphertexts. These are essential to the attack: the idea is to create and encrypt \emph{helper} images \( H_j \) of size \( M \times M \) such that we are able to tell exactly what shuffling and reversing has been done, and decrypt the \( \texttt{Enc}(H_j) \) at the same time as the blocks of \( \texttt{Enc}(I) \) to obtain the original image.

We choose to create \( M \times M = M^2 \) helper images for simplicity, although this number can be reduced by a constant as shown later. To create the helper images \( H_j \) of size \( M \times M \) with \( j \in [1, M^2] \), fix a pixel \( p^j \) unique to \( H_j \) (such that each pixel location in a block of size \( M \times M \) corresponds to a unique \( H_j \)). We choose four distinct integers \( a, b, c, z \in [0, 2^L - 1] \) such that \( a \), \( b \), \( c \), and \( z \) are distinct from \( (2^L - 1) - a \), \( (2^L - 1) - b \), \( (2^L - 1) - c \), and \( (2^L - 1) - z \). Then we set for \( p^j \) in \( H_j \) the color components \( p^j_R = a \), \( p^j_G = b \), and \( p^j_B = c \); we set for all other pixels \( q \) in \( H_j \) the color components \( q_R = q_G = q_B = z \). Uniqueness of the values ensures that we can sort the color components into their original locations, and the distinction between \( a \), \( b \), \( c \), and \( z \) against \( (2^L - 1) - a \), \( (2^L - 1) - b \), \( (2^L - 1) - c \), and \( (2^L - 1) - z \) ensures that we are able to tell when a value has been reversed.

After querying the encryption scheme with the helper image \( H_j \) to obtain its ciphertext \( \texttt{Enc}(H_j) \), we partially decrypt \( \texttt{Enc}(H_j) \) to recover the pixel \( p^j \) by undoing the shuffling and reversing. More precisely, to first undo the reversing we check for each pixel \( p \) the value of \( p_C \) with \( C \in \{ R, G, B \} \) to see if it is equal to \( (2^L - 1) - a \), \( (2^L - 1) - b \), or \( (2^L - 1) - c \), and if it is, subtract it from \( 2^L - 1 \) to obtain the original value. Then to undo the shuffling, we sort the color components of \( p^j \) into their original locations. We do this process for all \( H_j \) to recover the complete mapping between the color components in a block and the color components in its encryption. We decrypt \( \texttt{Enc}(I) \) to recover \( I \), as desired, by doing the same operations as above on each pixel \( p^P \) in each block \( P \) in \( \texttt{Enc}(I) \) corresponding to pixel \( p \) in \( \texttt{Enc}(H_j) \).

\begin{algorithm}[H]
\caption{{\bf Tanaka Chosen-Plaintext Attack Helper Image}.}
\KwIn{Block dimension \( M \);  number of bits \( L \); \( j \in [1, M^2] \).}
\KwOut{Helper image \( H_j \) of size \( M \times M \).} 
\vspace{0.1in}
\nl Instantiate empty image \( H_j \) of size \( M \times M \); \\
\nl Fix \( p^j = (u, v) \in H_j \) such that \( p^j \ne p^\ell \, \forall \, \ell \in [1, M^2] \setminus \{ j \} \); \\
\nl Choose distinct integers \( a, b, c, z \in [0, 2^L - 1] \) such that \( a \), \( b \), \( c \), \( d \) \\
\Indp are distinct from \( (2^L - 1) - a \), \( (2^L - 1) - b \), \( (2^L - 1) - c \), \( (2^L - 1) - z \); \\ \Indm
\nl \( p^j_R \gets a \); \\
\nl	\( p^j_G \gets b \); \\
\nl	\( p^j_B \gets c \); \\
\nl \ForEach{\( q = (u, v) \in H_j \)}{
\nl     \If{\( q \ne p^j \)}{
\nl		    \( q_R \gets z \); \\
\nl		    \( q_G \gets z \); \\
\nl		    \( q_B \gets z \);
        }
	}
\nl \Return{\( H_j \)};
\end{algorithm}

\newsavebox{\Enc}
\begin{algorithm}[H]
\caption{{\bf Tanaka Chosen-Plaintext Attack}.}
\begin{lrbox}{\Enc}
\verb=Enc=
\end{lrbox}
\KwIn{Encrypted input image \( \texttt{Enc}(I) \) of size \( U \times V \); number of bits \( L \); encrypted helper images \( \texttt{Enc}(H_j) \)'s of size \( M \times M \), chosen integers \( a, b, c \).}
\SetKwInOut{Modifies}{Modifies}
\Modifies{\( \texttt{Enc}(I) \), \( \texttt{Enc}(H_j) \)'s to decrypt them.} 
\vspace{0.1in}
\nl \For{\( j \gets 1 \) \KwTo \( M^2 \)}{
\nl     \ForEach{\( p = (u, v) \in \usebox{\Enc}(H) \)}{
\nl         \ForEach{\( C \in \{ R, G, B \} \)}{
\nl             \If{\( p_C = (2^L - 1) - a \vee p_C = (2^L - 1) - b \vee p_C = (2^L - 1) - c \)}{
\nl                 \( p_C \gets (2^L - 1) - p_C \); \\
\nl                 \ForEach(\tcp*[f]{For each MxM block P in Enc(I)}){\( P \in \usebox{\Enc}(I) \)}{
\nl                     \( p^P_C \gets (2^L - 1) - p^P_C \)\tcp*[r]{Change the corresponding pixel}
                    }
                }
            }
	    }
	}
\nl \For{\( j \gets 1 \) \KwTo \( M^2 \)}{
\nl     \( p \gets p^j = (u, v) \in \texttt{Enc}(H_j) \); \\
\nl     Sort color components within \( \texttt{Enc}(H_j) \) such that \( p_R = a \), \( p_G = b \), \( p_B = c \); \\
\nl     \ForEach{\( P \in \usebox{\Enc}(I) \)}{
\nl         Modify \( p^P_R, p^P_G, p^P_B \) in the same way as above; \\
        }
	}
\end{algorithm}

\begin{figure}[H]
\centering
\includegraphics[width = 96px, height = 96px]{Images/horse_enc}
\qquad
\includegraphics[width = 96px, height = 96px]{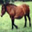}
\caption{An 8-bit image encrypted with the Tanaka scheme and its decryption using the attack above.}
\end{figure}

We described the attack above using 1-pixel subvidisions of the \( M \times M \) blocks, but we wish to consider larger subdivisions to reduce the number of helper images. Observe that the number of distinct integers in \( [0, 2^L - 1] \) that we can choose to assign to the color components is \( 2^L/2 = 2^{L - 1} \), because these integers must be distinct from their reverses as well. There are \( 3N \) color components in a subdivision with \( N \) pixels, so we must have \( 3N \le 2^{L - 1} \). This implies \( N \le 2^{L - 1}/3 \), so one can reduce the number of helper images by a factor of up to \( 2^{L - 1}/3 \) through undoing the shuffling and reversing of up to this many number of pixels in a block at once.

\subsection{Attacking the SKK Scheme}

For the SKK scheme we present separate attacks for the variants of the scheme using the same encryption key for all images verses different encryption keys for every image.

\subsubsection{Chosen-Plaintext Attack for Same Key Variant}

For the scheme using the same encryption key for each image, we present a full chosen-plaintext attack. One may also use the partial ciphertext-only attacks presented below when discussing different encryption keys.

Being able to query the encryption scheme once with a chosen plaintext to obtain the corresponding ciphertexts, we wish to fully decrypt a given ciphertext \( \texttt{Enc}(I) \) of the original 8-bit image \( I \). Similar to the attack against the Tanaka scheme, the idea is to create and encrypt a \emph{helper} image \( H \), this time of the same size as \( I \), such that we are able to tell exactly what negative-positive transforming and shuffling has been done, and decrypt \( \texttt{Enc}(H) \) at the same time as \( \texttt{Enc}(I) \) to obtain the original image.

To create the helper image \( H \) of the same size as \( I \), we choose three distinct integers \( a, b, c \in [0, 2^L - 1] \) such that \( a \), \( b \), and \( c \) are distinct from \( a \oplus (2^L - 1) \), \( b \oplus (2^L - 1) \), and \( c \oplus (2^L - 1) \). Then we set for each pixel \( p \) in \( H \) the color components \( p_R = a \), \( p_G = b \), and \( p_B = c \). Uniqueness of the values ensures that we can sort the color components into their original locations, and the distinction between \( a \), \( b \), and \( c \) against \( a \oplus (2^L - 1) \), \( b \oplus (2^L - 1) \), and \( c \oplus (2^L - 1) \) ensures that we are able to tell when a value has been negative-positive transformed.

After querying the encryption scheme with the helper image \( H \) to obtain its ciphertext \( \texttt{Enc}(H) \), we decrypt \( \texttt{Enc}(H) \) to recover \( H \) by undoing the negative-positive transforming and shuffling. More precisely, to first undo the negative-positive transforming we check for each pixel \( p \) the value of \( p_C \) with \( C \in \{ R, G, B \} \) to see if it is equal to \( a \oplus (2^L - 1) \), \( b \oplus (2^L - 1) \), or \( c \oplus (2^L - 1) \), and if it is, XOR it with \( 2^L - 1 \) to obtain the original value. Then to undo the shuffling, we sort the color components into their original locations. We decrypt \( \texttt{Enc}(I) \) to recover \( I \), as desired, by doing the same operations as above on each pixel \( p^I \) in \( \texttt{Enc}(I) \) corresponding to pixel \( p \) in \( \texttt{Enc}(H) \).

\begin{algorithm}[H]
\caption{{\bf SKK Chosen-Plaintext Attack Helper Image}.}
\KwIn{Input image dimensions \( U \times V \);  number of bits \( L \).}
\KwOut{Helper image \( H \) of size \( U \times V \).} 
\vspace{0.1in}
\nl Instantiate empty image \( H \) of size \( U \times V \); \\
\nl Choose distinct integers \( a, b, c \in [0, 2^L - 1] \) such that \( a \), \( b \), \( c \) \\
\Indp are distinct from \( a \oplus (2^L - 1) \), \( b \oplus (2^L - 1) \), \( c \oplus (2^L - 1) \); \\ \Indm
\nl \ForEach{\( p = (u, v) \in H \)}{
\nl		\( p_R \gets a \); \\
\nl		\( p_G \gets b \); \\
\nl		\( p_B \gets c \);
	}
\nl \Return{\( H \)};
\end{algorithm}

\begin{algorithm}[H]
\caption{{\bf SKK Chosen-Plaintext Attack}.}
\begin{lrbox}{\Enc}
\verb=Enc=
\end{lrbox}
\KwIn{Encrypted input image \( \texttt{Enc}(I) \) of size \( U \times V \); number of bits \( L \); encrypted helper image \( \texttt{Enc}(H) \) of size \( U \times V \), chosen integers \( a, b, c \).}
\SetKwInOut{Modifies}{Modifies}
\Modifies{\( \texttt{Enc}(I), \texttt{Enc}(H) \) to decrypt them.} 
\vspace{0.1in}
\nl \ForEach{\( p = (u, v) \in \usebox{\Enc}(H) \)}{
\nl     \ForEach{\( C \in \{ R, G, B \} \)}{
\nl         \If{\( p_C = a \oplus (2^L - 1) \vee p_C = b \oplus (2^L - 1) \vee p_C = c \oplus (2^L - 1) \)}{
\nl             \( p_C \gets p_C \oplus (2^L - 1) \); \\
\nl             \( p^I_C \gets p^I_C \oplus (2^L - 1) \); \\
            }
        }
\nl     Sort color components within pixel \( p \) in \( \texttt{Enc}(H) \) \\
\Indp such that \( p_R = a \), \( p_G = b \), \( p_B = c \); \\ \Indm
\nl     Modify \( p^I_R, p^I_G, p^I_B \) in the same way as above; \\
	}
\end{algorithm}

\begin{figure}[H]
\centering
\includegraphics[width = 96px, height = 96px]{Images/bird_enc}
\qquad
\includegraphics[width = 96px, height = 96px]{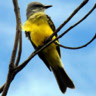}
\caption{An 8-bit image encrypted with the SKK scheme and its decryption using the attack above.}
\end{figure}

\subsubsection{Ciphertext Only Attacks for Different Keys Variant}

For the scheme using different encryption keys for each image, we present two \emph{partial} ciphertext-only attacks: that is, we are able to recover enough features of the original image to presumably be able to distinguish between two possible plaintexts the correct preimage of the ciphertext more often than at random.

\subparagraph{Basic Ciphertext-Only Attack.}

With access to ciphertexts only, we wish to partially decrypt a given ciphertext \( \texttt{Enc}(I) \) of the original \( L \)-bit image \( I \). To do this, we observe that the perception of edges in images is not determined so much by the specific color or ordered RGB coordinate triple of a pixel but by properties such as the combined magnitude of the color components of a pixel. Indeed, experimentation with permuting the color components of each pixel in some image reveals that the resulting image looks like a grayscale version of the original image; in particular, the objects are still identifiable.

\begin{figure}[H]
\centering
\includegraphics[width = 96px, height = 96px]{Images/bird}
\qquad
\includegraphics[width = 96px, height = 96px]{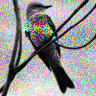}
\caption{An 8-bit image and the same image with its color components shuffled in each pixel. Note how the objects are still identifiable.}
\end{figure}

We also note that the gradient vector and gradient magnitude at each pixel are commonly used in computational edge detection, so properties such as the difference in magnitude of a color component value of a pixel with respect to the color component values of surrounding pixels also seem to be important. We observe that, for reasonably realistic images, areas between edges share similar or very gradually changing color component values. This implies that the gradient magnitude at each pixel is close to minimal.

We suspect the observations above are the reasons why the following partial attack works. First, we choose either 0 or 1 to be the leading bit of each color component. Then for each color component of each pixel in \( \texttt{Enc}(I) \), if its leading bit is not that number, we XOR the value with \( 2^L - 1 \) so that the leading bit is that number.

\begin{algorithm}[H]
\caption{{\bf SKK Basic Ciphertext-Only Attack}.}
\begin{lrbox}{\Enc}
\verb=Enc=
\end{lrbox}
\KwIn{Encrypted input image \( \texttt{Enc}(I) \) of size \( U \times V \); number of bits \( L \); leading bit \( b \in \{ 0, 1 \} \).}
\SetKwInOut{Modifies}{Modifies}
\Modifies{\( \texttt{Enc}(I) \) to partially decrypt it.} 
\vspace{0.1in}
\nl \ForEach{\( p = (u, v) \in \usebox{\Enc}(I) \)}{
\nl     \ForEach{\( C \in \{ R, G, B \} \)}{
\nl         \If{\( \lfloor p_C/2^{L - 1} \rfloor \ne b \)}{
\nl             \( p_C \gets p_C \oplus (2^L - 1) \); \\
            }
        }
	}
\end{algorithm}

\begin{figure}[H]
\centering
\includegraphics[width = 96px, height = 96px]{Images/bird_enc}
\qquad
\includegraphics[width = 96px, height = 96px]{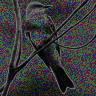}
\qquad
\includegraphics[width = 96px, height = 96px]{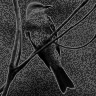}
\caption{An 8-bit image encrypted with the Sirichotedumrong, Kinoshita, and Kiya scheme, a possible decryption using the basic attack above, and a grayscale version of the decryption (this is sometimes clearer).}
\end{figure}

\begin{figure}[H]
\centering
\includegraphics[width = 96px, height = 96px]{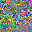}
\qquad
\includegraphics[width = 96px, height = 96px]{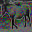}
\qquad
\includegraphics[width = 96px, height = 96px]{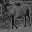}
\caption{Another 8-bit image encrypted with the Sirichotedumrong, Kinoshita, and Kiya scheme, a possible decryption using the basic attack above, and a grayscale version of the decryption.}
\end{figure}

Recall that the two main operations on the pixels in this encryption scheme are the negative-positive transformation and shuffle. Note that the shuffle operation does not change the combined magnitude of the color components of a pixel, so we essentially ignore this step. Then we concern ourselves with the negative-positive transformation; but it is likely in a reasonably realistic image that a color component value of a pixel has the same leading bit as those of surrounding pixels. So we change all color components of every pixel to have the same leading bit, and this likely restores the difference in magnitude of a color component value of a pixel with respect to the color component values of surrounding pixels and preserves areas between edges sharing similar or very gradually changing color component values. Hence, we get back an image with somewhat recognizable objects.

(It is easy to think of images that, when encrypted using the SKK scheme, this attack would fail against. Indeed, a pure black and white picture would be ``decrypted" as all-black or all-white by this algorithm. But such a picture would be non-realistic for the purposes of this attack.)

\subparagraph{More Advanced Ciphertext-Only Attack.}

With access to ciphertexts only, we wish to partially decrypt a given ciphertext \( \texttt{Enc}(I) \) of the original \( L \)-bit image \( I \). Let \( p \) be a pixel that we are trying to determine the original RGB coordinate triple of, and let \( q \) be a different nearby pixel for comparison. Then based off the discussion at the start of the basic attack above, one may conclude that minimizing the quantity
\begin{equation} \label{eqn:min}
\sum_{C \in \{ R, G, B \}} \lvert q_C - p_C \rvert
\end{equation}
would be an effective way to recover the original image \( I \), since this minimizes change in color component values. Indeed, this is the idea behind the more advanced attack: for each pixel, we try every possible shuffle permutation and negative-positive transformation selection on its encrypted color component values and pick the combination that minimizes the quantity above.

More precisely, let \( p \) in \( \texttt{Enc}(I) \) be a pixel that has not been decrypted yet, and let \( q \) in \( \texttt{Enc}(I) \) be a nearby pixel that has been decrypted. Consider the three color components \( p_R \), \( p_G \), and \( p_B \) of \( p \): there are \( 3! = 6 \) possible shuffle permutations and \( 2^3 = 8 \) possible negative-positive transformation selections, to obtain \( 6 \times 8 = 48 \) total possible options, at least one of which corresponds to the original state of the corresponding pixel in \( I \). We calculate the quantity above for each option and let the decrypted version of \( p \) be an option that minimizes that quantity. As for the first pixel \( p \), with no decrypted pixel \( q \) to compare to, one may fix the original state of this \( p \) from \( \texttt{Enc}(I) \) or generate 48 possible full-image decryptions by trying all possible options for the original state of the corresponding pixel in \( I \). 48 images is, of course, feasible to quickly scan by a human to find the most sensible decryption.

In the algorithm below, min\_diff\_opt instantiated in Line 3 stores the option of shuffling and negative-positive transforming that gives the minimal value so far in Lines 5-8 of (\ref{eqn:min}) and min\_diff instantiated in Line 4 stores that minimal value.

\begin{algorithm}[H]
\caption{{\bf SKK Advanced Ciphertext-Only Attack}.}
\begin{lrbox}{\Enc}
\verb=Enc=
\end{lrbox}
\KwIn{Encrypted input image \( \texttt{Enc}(I) \) of size \( U \times V \) with at least one decrypted pixel; number of bits \( L \).}
\SetKwInOut{Modifies}{Modifies}
\Modifies{\( \texttt{Enc}(I) \) to partially decrypt it.} 
\vspace{0.1in}
\nl \ForEach{\( p = (u, v) \in \usebox{\Enc}(I) \)}{
\nl     Choose nearby pixel that has been decrypted \( q \); \\
\nl     min\_diff\_opt \( \gets p \); \\
\nl     min\_diff \( \gets \sum_{C \in \{ R, G, B \}} \lvert q_C - p_C \rvert \); \\
\nl     \ForEach(\tcp*[f]{Each option is shuffle and neg-pos}){option \( p^* \) of \( p \)}{
\nl         \If{\( \sum_{C \in \{ R, G, B \}} \lvert q_C - p^*_C \rvert < \) min\_diff}{
\nl             min\_diff\_opt \( \gets p^* \); \\
\nl             min\_diff \( \gets \sum_{C \in \{ R, G, B \}} \lvert q_C - p^*_C \rvert \); \\
            }
        }
\nl \( p \gets \) min\_diff\_opt; \\
	}
\end{algorithm}

\begin{figure}[H]
\centering
\includegraphics[width = 96px, height = 96px]{Images/bird_enc}
\qquad
\includegraphics[width = 96px, height = 96px]{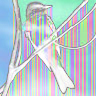}
\qquad
\includegraphics[width = 96px, height = 96px]{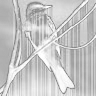}
\caption{An 8-bit image encrypted with the Sirichotedumrong, Kinoshita, and Kiya scheme, a possible decryption using the more advanced attack above, and a grayscale version of the decryption. Note how the left side of the decryption has coloring that reflects the coloring of the original image.}
\end{figure}

\begin{figure}[H]
\centering
\includegraphics[width = 96px, height = 96px]{Images/horse_enc_SKK}
\qquad
\includegraphics[width = 96px, height = 96px]{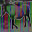}
\qquad
\includegraphics[width = 96px, height = 96px]{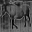}
\caption{Another 8-bit image encrypted with the Sirichotedumrong, Kinoshita, and Kiya scheme, a possible decryption using the more advanced attack above, and a grayscale version of the decryption.}
\end{figure}

Using this attack in experiments, we appear to sometimes get back images with objects more clear and recognizable than objects in the images from the basic attack, along with coloring that may partially reflect the coloring of the original image.

(Again, it is easy to think of non-realistic images that, when encrypted using the SKK scheme, this attack would fail against.)

\section{Conclusion}

Above, we presented a chosen-plaintext attack against the Tanaka scheme and both a chosen-plaintext attack and ciphertext-only attacks against the SKK scheme. As for the Tanaka scheme and SKK same key variant, we reiterate the fundamental weakness of deterministic encryption (i.e. that it can never be CPA secure) and point out that other techniques such as homomorphic encryption are usually probabilistic. As for the SKK different keys variant, more work needs to be done to develop image encryption schemes that presumably do not allow the gradient magnitude to be exploited in the same manner as above; but such schemes may deteriorate the performance of the encrypted images with machine learning and DNNs, so this may be an intrinsic trade-off between security and performance.

\bibliography{main}

\begin{thebibliography}{10}

\bibitem{boemer2019ngraph}
Fabian Boemer, Yixing Lao, Rosario Cammarota, and Casimir Wierzynski.
\newblock ngraph-he: a graph compiler for deep learning on homomorphically
  encrypted data.
\newblock {\em Proceedings of the 16th ACM International Conference on
  Computing Frontiers}, pages 3--13, 2019.

\bibitem{PrivateAI}
Benjamin~M. Case, Marcella Hastings, Siam Hussain, and Monika Trimoska.
\newblock Happykidz: Privacy preserving phone usage tracking, 2020.

\bibitem{gilad2016cryptonets}
Nathan Dowlin, Ran Gilad-Bachrach, Kim Laine, Kristin Lauter, Michael Naehrig,
  and John Wernsing.
\newblock Cryptonets: Applying neural networks to encrypted data with high
  throughput and accuracy.
\newblock {\em International Conference on Machine Learning}, pages 201--210,
  2016.

\bibitem{KatzLindell}
Jonathan Katz and Yehuda Lindell.
\newblock {\em Introduction to Modern Cryptography}, page~72.
\newblock CRC Press, second edition, 2015.

\bibitem{lou2019glyph}
Qian Lou, Bo~Feng, Geoffrey~C Fox, and Lei Jiang.
\newblock Glyph: Fast and accurately training deep neural networks on encrypted
  data.
\newblock {\em arXiv:1911.07101}, 2019.

\bibitem{Madono}
Koki Madono, Masayuki Tanaka, Masaki Onishi, and Tetsuji Ogawa.
\newblock Block-wise scrambled image recognition using adaptation network.
\newblock In {\em AAAI WS}, 2020.

\bibitem{privacypreservingcomp}
Privacy-preserving computation techniques.
\newblock http://publications.officialstatistics.org/
  handbooks/privacy-preserving-techniques-handbook/UN\%20Handbook\%20for\%20
  Privacy-Preserving\%20Techniques.pdf, 2019.

\bibitem{SKK}
Warit Sirichotedumrong, Yuma Kinoshita, and Hitoshi Kiya.
\newblock Pixel-based image encryption without key management for
  privacy-preserving deep neural networks.
\newblock {\em IEEE Access}, 7:177844--177855, 2019.

\bibitem{SMKK}
Warit Sirichotedumrong, Takahiro Maekawa, Yuma Kinoshita, and Hitoshi Kiya.
\newblock Privacy-preserving deep neural networks with pixel-based image
  encryption considering data augmentation in the encrypted domain.
\newblock {\em 2019 IEEE International Conference on Image Processing (ICIP)},
  pages 674--678, 2019.

\bibitem{TanakaPublished}
Masayuki Tanaka.
\newblock Learnable image encryption.
\newblock {\em 2018 IEEE International Conference on Consumer
  Electronics-Taiwan (ICCE-TW)}, pages 1--2, 2018.

\end{thebibliography}
\bibliographystyle{plain}

\end{document}